\documentclass[aps,
               prb,
               twocolumn,
               10pt,
               nofootinbib,
               superscriptaddress,
               amsfonts,
               amsmath,
               amssymb]{revtex4-2}

\usepackage{bm}
\usepackage{siunitx}
\usepackage{microtype}
\usepackage{upgreek}
\usepackage{graphicx}
\usepackage{hyperref}
\usepackage{xcolor}

\usepackage{physics}

\hypersetup{
    colorlinks,
    linkcolor={blue!80!black},
    citecolor={blue!80!black},
    urlcolor={blue!80!black}
}

\begin{document}

\title{Directional conductance of Andreev crystals in hybrid Josephson junction arrays}

\author{Anders~Enevold~Dahl}
\affiliation{Center for Quantum Devices, Niels Bohr Institute, University of Copenhagen, DK-2100 Copenhagen, Denmark}
\author{Andrea~Maiani}
\affiliation{Center for Quantum Information Physics, Department of Physics, New York University, New York, NY 10003 USA}
\affiliation{Nordita, KTH Royal Institute of Technology and Stockholm University, Hannes Alfvéns väg 12, SE-10691 Stockholm, Sweden}
\author{Max~Geier}
\affiliation{Department of Physics, Massachusetts Institute of Technology, Cambridge, MA 02139, USA}
\author{Javad~Shabani}
\affiliation{Center for Quantum Information Physics, Department of Physics, New York University, New York, NY 10003 USA}
\author{Karsten~Flensberg}
\affiliation{Center for Quantum Devices, Niels Bohr Institute, University of Copenhagen, DK-2100 Copenhagen, Denmark}
\date{\today}

\begin{abstract}
Andreev bound states are coherent electron-hole superpositions that form in a normal metal through repeated Andreev reflection at a superconducting interface.
When the length of a superconducting segment is comparable to the coherence length, the bound states on opposite sides of the segment hybridize through quasiparticle tunneling.
In a periodic array, these hybridized Andreev bound states form energy bands below the superconducting gap.
We develop a theoretical framework for transport in such Andreev crystals.
We demonstrate that, at high interface transparency, a constant phase bias between neighboring superconductors renders the bands \emph{directional}: one band contains only right-moving and the other only left-moving electronic states.
This property leads to a directional conductance that enables the device to operate as a flux- and bias-voltage-tunable filter that allows signal transmission in only one direction.
\end{abstract}

\maketitle

%%%%%%%%%%%%%%%%%%%%%%%%%%%%%%%%%%%%%%%%
% INTRODUCTION
%%%%%%%%%%%%%%%%%%%%%%%%%%%%%%%%%%%%%%%%
\section{Introduction}
\label{sec:introduction}

Andreev reflection is a fundamental process in hybrid heterostructures. It occurs at the interface between a superconductor and a metal, where an incident electron from the metal is retro-reflected as a hole, or vice versa, with the corresponding transfer of a Cooper pair into or out of the superconducting condensate~\cite{blonderTransitionMetallicTunneling1982}. In systems with multiple superconducting interfaces, consecutive Andreev reflections can interfere constructively and generate Andreev bound states (ABSs), localized subgap states whose energy depends on the relative phases of the superconducting leads~\cite{beenakkerUniversalLimitCriticalcurrent1991, beenakkerSUPERCONDUCTINGQUANTUMPOINT1992a, johannsenFermionicQuantumSimulation2025}. The ABSs mediate the transfer of Cooper pairs across the junctions, thereby establishing a direct relation between the ABS spectrum and the current-phase relation. This relation makes it possible to engineer Josephson junctions with tailored current-phase relations by designing and controlling the ABS spectrum, opening new possibilities for tailored superconducting devices for quantum technologies~\cite{larsenSemiconductorNanowireBasedSuperconductingQubit2015, casparisSuperconductingGatemonQubit2018, aguadoPerspectiveSemiconductorbasedSuperconducting2020, banerjeeControlAndreevBound2023, banerjeeSignaturesTopologicalPhase2023, stricklandGatemoniumVoltageTunableFluxonium2024, coraiolaPhaseengineeringAndreevBand2023}.

In recent years, the development of hybrid superconductor-semiconductor structures has substantially advanced the field of superconducting electronics~\cite{Shabani2016, ingla-aynesEfficientSuperconductingDiodes2025}. These heterostructures incorporate semiconductor nanostructures in contact with superconductors, allowing the creation of a proximitized low-dimensional electron gas with tunable density and spin-orbit coupling~\cite{moehleInSbAsTwoDimensionalElectron2021}. One remarkable feature of such heterostructures is their long coherence length, which originates from the low Fermi velocity in the semiconductor. The long coherence length strongly enhances crossed Andreev reflection~\cite{LeijnseCouplingSpinQubits2013}, a process in which an electron incident at one superconductor interface is converted into a hole at a spatially separated interface. This process is central to applications such as Cooper-pair splitters~\cite{hofstetterCooperPairSplitter2009, hofstetterFiniteBiasCooperPair2011, schindeleNearUnityCooperPair2012, herrmannCarbonNanotubesCooperPair2010} and minimal Kitaev chains~\cite{leijnseParityQubitsPoor2012, sauRobustnessMajoranaFermions2010, fulgaAdaptiveTuningMajorana2013, dvirRealizationMinimalKitaev2023, tenhaafTwositeKitaevChain2024, escribanoPhasecontrolledMinimalKitaev2025a}.

\begin{figure}[ht]
    \centering
    \includegraphics[width=1\linewidth]{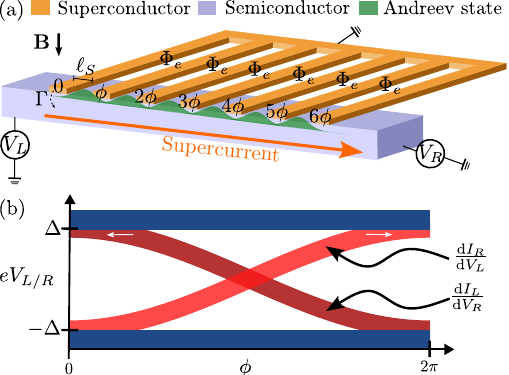}
    \caption{
    (a)
    Design of an Andreev crystal: sketch of a Josephson junction array with seven superconductors in contact with a semiconducting nanowire. Neighboring superconductors differ in phase by $\phi$.
    The phase-biased Josephson junctions host a supercurrent that flows along the array.
    (b) Illustration of the differential conductance as a function of phase and voltage bias.
    The two bands (dark and light red) represent the conducting regions in the nonlocal conductance: at high junction transparency and finite phase bias $\phi \neq 0, \pi$, each band consists only of left- or right-movers, respectively (indicated by white arrows). This leads to directional conductance: electrons tunneling into the array from the right (left) can transmit through the array only within the band consisting of left- (right-) movers.
    }
    \label{fig:Fig1}
\end{figure}

Another consequence of crossed Andreev reflection in hybrid heterostructures is the possibility that ABSs in adjacent semiconductor junctions hybridize, leading to so-called Andreev molecules~\cite{pilletNonlocalJosephsonEffect2019, pilletScatteringDescriptionAndreev2020, haxellDemonstrationNonlocalJosephson2023, matsuoPhasedependentAndreevMolecules2023, matsuoPhaseEngineeringAnomalous2023}. Andreev molecules represent an interesting regime in which the ABSs extend over multiple junctions, thereby creating coherent coupling between distant superconductors.

Building on this concept, we propose a further step: periodic nanostructures in which superconducting segments are arranged in a regular lattice. In such a system, the hybridization of ABSs can occur not only between adjacent junctions but also across an entire array of superconducting segments. The result is the formation of Andreev bands, analogous to Bloch bands in crystalline solids, forming what we define as \emph{Andreev crystals}~\cite{roucoSpectralPropertiesAndreev2021, roucoGapInversionQuasionedimensional2021}. The hybridization of ABSs in adjacent hybrid junctions is the defining feature of the Andreev crystal and distinguishes this structure from conventional Josephson junction arrays.

Short Andreev crystals were studied experimentally in the linear-response regime in Ref.~\cite{lykkegaardPhaseBiasedAndreevDiffraction2025}. Here we investigate the finite-bias transport properties that allow direct observation of the tunable left- and right-movers bands. We develop a theoretical framework for Andreev crystals and explore key phenomena such as nonlocal transport, which arises from coherent coupling between distant superconducting segments and enables nonlocal Josephson effects.

As a concrete application, we show that an Andreev crystal with high-transparency junctions can operate as a directional signal filter between voltage-biased metallic leads by exploiting the fact that left- and right-moving electrons give rise to different branches of the ABS spectrum. When different dc bias voltages are applied to the left and right terminals, charge transmission becomes possible in one direction while being strongly suppressed in the opposite direction. This directional transport forms the basis for the directional filter functionality discussed in the final section, where we outline potential device applications.

%%%%%%%%%%%%%%%%%%%%%%%%%%%%%%%%%%%%%%%%
% MODEL
%%%%%%%%%%%%%%%%%%%%%%%%%%%%%%%%%%%%%%%%
\section{Model}
\label{sec:model}

We consider a semiconducting nanowire partially covered by an array of superconducting finger leads, as illustrated in Fig.~\ref{fig:Fig1}. Each superconducting segment is characterized by a macroscopic phase $\phi_i$, where $i = 0, 1, 2, \dots$ is the lead index, and induces a local pairing potential $\Delta_i = \Delta e^{i\phi_i}$ in the electron gas through the proximity effect.

Neighboring superconductors are phase biased such that the phase difference
\begin{equation}
    \phi_{i+1} - \phi_i = \phi
    \label{eq:uniform-phase-difference}
\end{equation}
is constant along the array. This configuration realizes a \emph{discrete phase gradient}: the phase varies only across the intervening normal regions where the condensate is absent, in contrast to a continuous supercurrent-carrying state characterized by a uniform phase derivative $\partial_x \phi \neq 0$.
In this regime, Josephson coupling between adjacent superconductors transfers Cooper pairs across the normal segments, leading to coherent hybridization of Andreev bound states throughout the array, while the discrete phase gradient acts as an effective static gauge field for quasiparticle transport across the hybrid array.

Experimentally, the phase bias $\phi$ can be controlled in two equivalent ways.
One approach is to connect consecutive superconducting fingers through individual superconducting loops threaded by a magnetic flux $\Phi$, for which $\phi = 2\pi \Phi / \Phi_0$, with $\Phi_0 = h/2e$ being the flux quantum, as shown in Fig.~\ref{fig:Fig1}(a).
Alternatively, a single superconducting meander line carrying a steady supercurrent can impose the same uniform phase increment between neighboring contacts~\cite{lykkegaardPhaseBiasedAndreevDiffraction2025}.
The resulting chain of phase-biased junctions constitutes a one-dimensional hybrid Josephson lattice, which is the minimal realization of an Andreev crystal.

\subsection{Scattering matrix approach}
\label{sec:scattering-matrix-approach}

We use a scattering-matrix approach to model the hybrid Josephson junction array sketched in Fig.~\ref{fig:Fig1}(a). The model is constructed by successive concatenation of normal-superconductor-normal (NSN) segments. We start from the clean case, in which an electron can undergo only Andreev reflection or transmission through the superconductor, and later include backscattering in the normal regions. The scattering matrix for a superconducting segment takes the form
\begin{align}
\label{eq:superconducting-scattering-matrix}
    S_S(\omega) = \mqty(
    0 & r_{eh}(\omega) & t_{ee}(\omega) & 0 \\
    r_{he}(\omega) & 0 & 0 & t_{hh}(\omega) \\
    t_{ee}(\omega) & 0 & 0 & r_{eh}(\omega) \\
    0 & t_{hh}(\omega) & r_{he}(\omega) & 0).
\end{align}
Here $\omega$ denotes the quasiparticle energy, $r_{he}$ ($r_{eh}$) is the amplitude for an electron (hole) to undergo Andreev retroreflection as a hole (electron), and $t_{ee}$ ($t_{hh}$) is the amplitude for an electron (hole) to be transmitted through the superconducting segment.

The scattering matrix obeys particle-hole symmetry,
\begin{equation}
    \mathcal{P} S(\omega) \mathcal{P}^{-1} = S^*(-\omega),
    \label{eq:particle-hole-symmetry}
\end{equation}
where the particle-hole symmetry operator is $\mathcal{P} = i\tau_y \mathcal{K}$.
Here, $\mathcal{K}$ denotes complex conjugation and $\{\tau_i\}$ are the Pauli matrices acting in Nambu space.
This convention corresponds to symmetry class~D, for which $\mathcal{P}^2 = +1$.
The symmetry constraint reduces the number of independent scattering amplitudes, since
\begin{align}
    r_{eh}(\omega) &= -r^*_{he}(-\omega),
    \label{eq:particle-hole-r-eh}
    \\
    t_{hh}(\omega) &= t^*_{ee}(-\omega),
    \label{eq:particle-hole-t-hh}
\end{align}
as discussed in Ref.~\cite{maianiConductanceMatrixSymmetries2022}.

We match the Bogoliubov--de Gennes wave functions at the normal--superconductor interfaces.
Within the Andreev approximation, obtained by linearizing the dispersion around the Fermi energy, we find
\begin{subequations}
\label{eq:nsn-amplitudes}
\begin{align}
    r_{he} &= \frac{1 - t_\mathrm{S}^2(x)}{1 - r_\Delta^2(x)t_\mathrm{S}^2(x)}\,r_\Delta(x)\, e^{-i\phi},
    \label{eq:r-he-amplitude}
    \\
    t_{ee} &= \frac{1 - r_\Delta^2(x)}{1 - r_\Delta^2(x)t_\mathrm{S}^2(x)}\, t_\mathrm{S}(x)\,e^{ik_{\mathrm{F}}\ell_{S}},
    \label{eq:t-ee-amplitude}
\end{align}
\end{subequations}
where $\ell_S$ is the width of the superconducting segment and $k_{\mathrm{F}}$ is the Fermi wave number in that segment; see Appendix~\ref{sec:appendix-nsn-scattering-matrix} for details.
We define $r_\Delta$ as the bare Andreev-reflection amplitude and $t_\mathrm{S}$ as the bare electron cotunneling amplitude.
Both functions depend on the dimensionless energy $x \equiv \omega/\Delta$.

The bare Andreev-reflection amplitude is
\begin{equation}
    r_\Delta(x) \equiv e^{-i\arccos(x)},
    \label{eq:bare-andreev-reflection}
\end{equation}
which is the phase acquired upon reflection at a normal-metal--semi-infinite-superconductor interface~\cite{beenakkerJosephsonCurrentSuperconducting1991}. The bare electron cotunneling amplitude is
\begin{align}
    t_\mathrm{S}(x) &\equiv e^{iq(x)\ell_{S}},
    \label{eq:bare-electron-cotunneling}
    \\
    q^{-1}(x) &\equiv \frac{\pi\xi_0}{\sqrt{x^2 - 1}},
    \label{eq:bare-penetration-length}
\end{align}
where $q^{-1}$ is the penetration length, an energy-dependent generalization of the BCS coherence length $\xi_0 \equiv \hbar v_\mathrm{F}/(\pi \Delta)$, with $v_\mathrm{F}$ as the Fermi velocity.

\subsection{Imperfect transmission in the semiconductor}
\label{sec:imperfect-transmission-semiconductor}

We incorporate charge-conserving backscattering in the normal semiconducting junction segments through a scattering matrix of the form
\begin{equation}
    S_N =
    \begin{pmatrix}
        r & 0 & t & 0 \\
        0 & r & 0 & t \\
        t & 0 & -r & 0 \\
        0 & t & 0 & -r
    \end{pmatrix},
    \label{eq:normal-scattering-matrix}
\end{equation}
where $r$ and $t$ denote, respectively, the reflection and transmission amplitudes for quasiparticles in the normal region. In the short-junction limit, these amplitudes can be taken as energy independent and parametrized by a single real coefficient, the junction transparency $T \equiv |t|^2 \in [0,1]$, such that
\begin{equation}
    t = \sqrt{T},
    \qquad
    r = \sqrt{1-T}.
    \label{eq:transparency-parametrization}
\end{equation}
The transparency $T$ represents the probability of a quasiparticle being transmitted across the normal segment.

The full device scattering matrix is then obtained by concatenating the normal and superconducting segments using the Redheffer star product, as detailed in Appendix~\ref{sec:appendix-concatenation}.
This construction assumes the short-junction limit, so that the phase accumulated during propagation through the semiconductor can be neglected.
For simplicity, we also neglect backscattering within the superconducting segments, such that all disorder effects are effectively captured by the single phenomenological parameter $T$.

\subsection{Proximity self-energy}
\label{sec:proximity-self-energy}

In the above expressions, the superconducting segments are described by an effective energy-independent pairing potential $\Delta$. In the proposed hybrid device, however, the superconducting correlations in the semiconductor are proximity-induced by the superconducting leads, as shown in Fig.~\ref{fig:Fig1}(a). In this setup, virtual tunneling in and out of the parent superconductor is more accurately described by a self-energy term of the form~\cite{stanescuProximityEffectSuperconductortopological2010, sauRobustnessMajoranaFermions2010, danonInteractionEffectsProximityinduced2015, hansenPhasetunableMajoranaBound2016}
\begin{equation}
    \Sigma(\omega, \Delta_0) = \Gamma \frac{-\omega + \Delta_0 \tau_x}{\sqrt{\Delta_0^2 - \omega^2}},
    \label{eq:self-energy}
\end{equation}
where $\Gamma$ is the coupling strength between the superconductor and semiconductor, proportional to the square of the tunneling amplitude times the normal density of states in the superconducting shell, and $\Delta_0$ is the gap of the parent superconductor.

By introducing the self-energy into the system's Green function as
\begin{align}
    G^{-1}(\omega) = \omega - H_0 - \Sigma(\omega,\Delta_0),
    \label{eq:greens-function}
\end{align}
we can verify how the coupling to the superconducting segments  modifies the quasiparticle spectrum in the semiconductor by a renormalized excitation energy and an effective gap described by
\begin{align}
    E(\omega) &= \omega \left( 1 + \frac{\Gamma}{\sqrt{\Delta_0^2 - \omega^2}} \right),
    \label{eq:renormalized-energy}
    \\
    \Delta(\omega) &= \frac{\Delta_0 \Gamma}{\sqrt{\Delta_0^2 - \omega^2}}.
    \label{eq:renormalized-gap}
\end{align}

The renormalization and the energy-dependent gap parameter in Eqs.~(\ref{eq:renormalized-energy}-\ref{eq:renormalized-gap}) alter the dimensionless ratio between energy and the gap, which can be incorporated into the results of the previous subsection by the replacement
\begin{align}
x \rightarrow x(\omega) = \frac{E(\omega)}{\Delta(\omega)} = \frac{\omega}{\Delta_0} \left(1 + \frac{\Delta_0}{\Gamma} \sqrt{1 - \frac{\omega^2}{\Delta_0^2}} \right).
\label{eq:x-replacement}
\end{align}
The proximity-induced renormalization increases the effective gap at low energies, where the induced pairing is controlled by $\Gamma$, while the correction vanishes as $\omega$ approaches $\Delta_0$.

An important consequence of this energy dependence is the renormalization of the quasiparticle penetration length $q^{-1}$ into the superconducting region. Since the coherence length is tied to the effective induced gap, it also becomes energy dependent. The renormalized decay length is
\begin{align}
    q^{-1} \rightarrow q^{-1}(\omega) = -i \pi \frac{\hbar v_F}{\pi \Delta(\omega)}\frac{1}{\sqrt{1 - x^2(\omega)}},
    \label{eq:renormalized-penetration-length}
\end{align}
which simplifies in the low-energy limit to
\begin{align}
    q^{-1} &\xrightarrow{\omega \rightarrow 0} -i\pi\xi_\Gamma,
    \quad\text{with}\quad
    \xi_\Gamma \equiv \frac{\hbar v_F}{\pi\Gamma}.
\label{eq:low-energy-penetration-length}
\end{align}

The renormalized energy ratio in Eq.~\eqref{eq:x-replacement}, $x(\omega)=E(\omega)/\Delta(\omega)$, exceeds unity beyond the induced minigap edge in the semiconductor. In this regime, $q(\omega)$ becomes real, and the corresponding decay length $q^{-1}$ ceases to describe evanescent modes, signaling the onset of continuum states with propagating plane-wave character. The transition between evanescent and propagating behavior, which determines the effective boundary of the induced gap, is further discussed in Sec.~\ref{sec:andreev-crystal} in the context of the Andreev crystal.

\begin{figure}[htb]
    \centering
    \includegraphics[width=0.9\columnwidth]{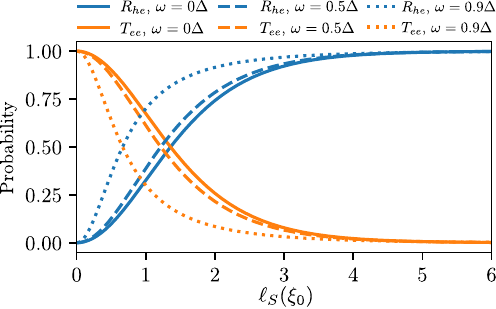}
    \caption{Transmission probability, $T_{ee}=|t_{ee}|^2$, and Andreev-reflection probability, $R_{he}=|r_{he}|^2$, of a single superconducting segment as a function of the segment length for three representative excitation energies. The plots are generated using Eqs.~\eqref{eq:nsn-amplitudes} and~\eqref{eq:renormalized-penetration-length}.}
    \label{fig:NSN_scattering}
\end{figure}

Fig.~\ref{fig:NSN_scattering} displays the normal-transmission and Andreev-reflection probabilities, $T_{ee}=|t_{ee}|^{2}$ and $R_{he}=|r_{he}|^{2}$, respectively, obtained from Eq.~\eqref{eq:nsn-amplitudes} with the renormalized decay length in Eq.~\eqref{eq:renormalized-penetration-length} as a function of the superconducting segment length $\ell_{S}$ for three representative excitation energies. For ultrashort segments, $\ell_{S}\ll\xi_0$, the quasiparticle traverses the proximitized region almost unimpeded, yielding $T_{ee}\approx1$ and $R_{he}\approx0$. As $\ell_{S}$ becomes comparable to the coherence length, the electron acquires an increasing hole component, and the probability of Andreev retroreflection grows at the expense of direct transmission. In the long-segment limit, $\ell_{S}\gg\xi_0$, the quasiparticle is fully Andreev reflected, so $R_{he}\to1$.

\subsection{Differential conductance}
\label{sec:differential-conductance}

We calculate the differential conductance matrix 
$G_{\alpha\beta} = \tfrac{\dd I_\alpha}{\dd V_\beta}$ 
using the Landauer--Büttiker approach; see Appendix~\ref{sec:appendix-landauer-buttiker} for details.
The zero-temperature differential conductance is given by
\begin{subequations}
\label{eq:zero-temperature-conductance}
\begin{align}
    G_{\alpha\alpha}^{(0)}(\omega) &=
    G_0 \left[ N_\alpha - |r^{\alpha\alpha}_{ee}|^2 + |r^{\alpha\alpha}_{he}|^2 \right],
    \label{eq:zero-temperature-local-conductance}
    \\
    G_{\alpha\beta}^{(0)}(\omega) &=
    -G_0 \left[ |t_{ee}^{\alpha\beta}|^2 - |t_{he}^{\alpha\beta}|^2 \right],
    \label{eq:zero-temperature-nonlocal-conductance}
\end{align}
\end{subequations}
where $G_0 = 2e^2/h$ is the conductance quantum and $N_\alpha$ is the number of channels in lead $\alpha$~\cite{takaneConductanceFormulaMesoscopic1992, danonNonlocalConductanceSpectroscopy2020}.
The finite-temperature conductance follows from thermal averaging as
\begin{align}
    G_{\alpha\beta}(e V_\beta) =
    \int_{-\infty}^{\infty} \dd\omega\,
    G_{\alpha\beta}^{(0)}(\omega - e V_\beta)\, [-f'(\omega)],
    \label{eq:conductance-finite-T}
\end{align}
with
\begin{equation}
[-f'(\omega)] =
\frac{1}{2k_B \theta}\,
\frac{1}{1 + \cosh(\omega/k_B \theta)},
\label{eq:thermal-kernel}
\end{equation}
which is the negative derivative of the Fermi--Dirac distribution, where $\theta$ denotes the electron temperature and $k_B$ is the Boltzmann constant.

%%%%%%%%%%%%%%%%%%%%%%%%%%%%%%%%%%%%%%%%%%%
% RESULTS
%%%%%%%%%%%%%%%%%%%%%%%%%%%%%%%%%%%%%%%%%%%
\section{Transport properties}
\label{sec:transport-properties}

To characterize the behavior of the Andreev crystal under finite bias, we now examine its transport response, starting from the elementary building block and then moving to extended arrays.

%%%%%%%%%%%%%%%%%%%%%%%%%%%%% NSNSN case %%%%%%%%%%%%%%%%%%%%%%%%%%%%%
\subsection{NSNSN junction}
\label{sec:nsnsn-junction}

We start with the simplest case: an NSNSN device, \textit{i.e.}, a single Josephson junction connected to two normal leads, at perfect transparency $T = 1$.

For energies below the gap, $\omega < \Delta$, the scattering matrix can be obtained analytically, and it has the same form as the superconducting scattering matrix in Eq.~\eqref{eq:superconducting-scattering-matrix}, with coefficients
\begin{widetext}
\begin{subequations}
\label{eq:nsnsn-analytic-amplitudes}
\begin{align}
    r_{he,LL}^{(1)}(x, \phi) &= \frac{r_\Delta(x)(t_\mathrm{S}^2 - 1)}{r_\Delta(x)^2 t_\mathrm{S}^2 - 1} \left[1 + \frac{t_\mathrm{S}^2 (r_\Delta^2(x) - 1)(r_\Delta^{*2}(-x) - 1) }{r_\Delta(x)r_\Delta^*(-x) (t_\mathrm{S}^2 - 1)^2 + e^{i\phi} (r_\Delta^2(x) t_\mathrm{S}^2 - 1) (r_\Delta^{*2}(-x) t_\mathrm{S}^2 - 1)} \right],
    \label{eq:nsnsn-r-he-ll}
    \\
    t_{ee,LR}^{(1)}(x, \phi) &=
    \frac{ t^2_\mathrm{S} (r^2_\Delta(x) - 1)^2 }{(1 - r^2_\Delta(x)t^2_\mathrm{S})^2 } \left[1 + \frac{e^{-i \phi} r_\Delta(x)r_\Delta^*(-x) (1 - t^2_\mathrm{S})^2 }{(1 - r^2_\Delta(x) t^2_S) (1 - r^{*2}_\Delta(-x) t^2_S) }\right]^{-1} e^{2ik_{\mathrm{F}}\ell_{S}},
    \label{eq:nsnsn-t-ee-lr}
\end{align}
\end{subequations}
\end{widetext}
where we used $t_\mathrm{S} = t_\mathrm{S}(x) = t_{\mathrm{S}}(-x)$. Here the superscript $(1)$ labels the single-junction NSNSN building block, namely one Josephson junction between the two normal terminals, and distinguishes it from the multi-junction array treated below. Analogous expressions hold for $r_{he,RR}^{(1)}$ and $t_{ee,RL}^{(1)}$, with $r_{he,RR}^{(1)}(x,\phi) = r_{he,LL}^{(1)}(x,-\phi) e^{-i\phi}$ and $t_{ee,RL}^{(1)}(x,\phi) = t_{ee,LR}^{(1)}(x,-\phi)$.

\begin{figure*}[ht]
    \centering
    \includegraphics{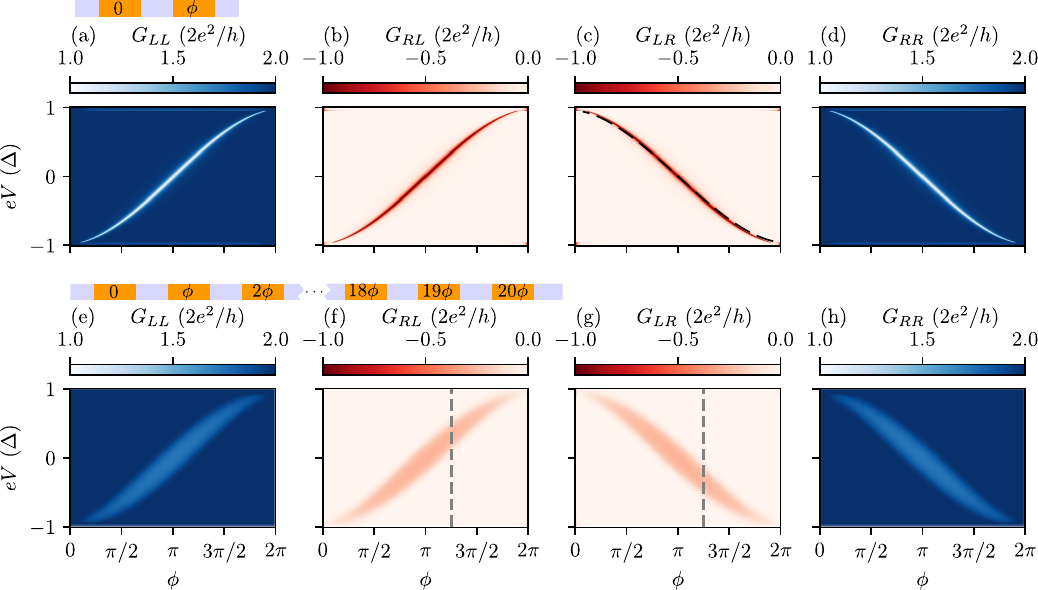}
    \caption{Top panels (a--d): Differential conductance, $G_{\alpha\beta}$, of an NSNSN device with phase difference $\phi$ between the superconductors and bias $eV$. A sketch of the setup is shown at the top left. Panels (a) and (d) display the local conductance, $\alpha = \beta$, and panels (b) and (c) the nonlocal conductance, $\alpha \neq \beta$. The dashed line in panel (c) follows from the resonance condition $\pi = 2\arccos[x(eV)]$, discussed in the text. Bottom panels (e--h): Differential conductance of a device with 21 superconductors and phase difference $\phi$ between neighboring superconductors. A sketch of the setup is again shown at the top left. Panels (e) and (h) display the local conductance, $\alpha = \beta$, and panels (f) and (g) the nonlocal conductance, $\alpha \neq \beta$. In panels (f) and (g), the dashed gray lines mark the regime discussed further in Fig.~\ref{fig:G1d}. In all simulations, $\Delta_0 = \SI{340}{\mu eV}$, $m^* = 0.023m_e$, $\Gamma = 6\Delta_0$, $\ell_S = 3\xi_0$, $\mu = \SI{1}{meV}$, and $\theta = \SI{30}{mK}$.}
    \label{fig:GLR}
\end{figure*}

The conductance matrix for the NSNSN case is shown in the top row of Fig.~\ref{fig:GLR}. Both the local and nonlocal conductances exhibit a peak that shifts with the relative phase $\phi$. Importantly, the conductances have opposite slopes depending on which lead is biased, leading to the directional asymmetry induced by the phase difference. This can be seen from Eq.~\eqref{eq:nsnsn-t-ee-lr}, where the phase bias enters through the factor $e^{-i\phi}r_\Delta(x)r_\Delta^*(-x)$, whose complex phase is $\phi-\pi+2 \arccos(x)$. Using $x(\omega)$ from Eq.~\eqref{eq:x-replacement}, a resonance condition arises at an energy that depends on the phase bias.
When this phase equals $\pi$, corresponding to $x=\cos(\phi/2)$, the transmission in Eq.~\eqref{eq:nsnsn-t-ee-lr} is maximal. Indeed, the subgap nonlocal conductance in Fig.~\ref{fig:GLR}(c) closely follows this dependence, shown there as the dashed line. The left- and right-moving states are thus separated in energy.

To better understand the resonant behavior, we expand the nonlocal differential conductance at low bias and for $\phi \simeq \pi$, where it acquires an approximate Lorentzian line shape with a broadening that depends on the superconducting-segment length. In particular, we find
\begin{equation}
        G_{LR}(0,\phi) \approx -G_0\frac{\lambda^2_\phi(\ell_{S})}{\left( \phi - \pi \right)^2 + \lambda_\phi^2(\ell_{S})},
\label{eq:nsnsn-lorentzian-phase}
\end{equation}
and
\begin{align}
    G_{LR}(x,\pi) &\approx -G_0\frac{\lambda_x^2(\ell_{S})}{x^2 + \lambda_x^2(\ell_{S})},
\label{eq:nsnsn-lorentzian-energy}
\end{align}
where $\lambda_\phi(\ell_{S}) = 2\csch\left(\tfrac{2\ell_{S}}{\xi_0}\right)$ and $\lambda_x(\ell_{S}) = \csch^2\left(\tfrac{\ell_{S}}{\xi_0}\right)$ are the half-widths at half maximum of the conductance features. It is evident that, for long superconducting segments, the conductance vanishes exponentially because the Andreev bound states do not extend deeply into the leads.

%%%%%%%%%%%%%%%%%%%%%%%%%%%%% Andreev crystal case %%%%%%%%%%%%%%%%%%%%%%%%%%%%%
\subsection{Andreev crystal}
\label{sec:andreev-crystal}

Next, we consider an extended array. Let $S^{(i)}$ denote the scattering matrix of the truncated array containing superconducting segments with indices $0,1,\dots,i$, with corresponding phases $\phi_0,\dots,\phi_i$, such that $i=0$ corresponds to the first segment. The recursion below updates the amplitudes when segment $i+1$ is added. Assuming perfect transmission throughout, we obtain the following recursion relations for the scattering amplitudes:
\begin{subequations}
\label{eq:andreev-crystal-recursions}
\begin{align}
    r_{\mathrm{he},LL}^{(i+1)} &= r_{\mathrm{he},LL}^{(i)} + \frac{r_{\mathrm{he}}^{(0)} e^{-i\phi_{i+1}} t_{\mathrm{ee},LR}^{(i)} t_{\mathrm{hh},LR}^{(i)}}{1 - r_{\mathrm{eh},LL}^{(i)} r_{\mathrm{he}}^{(0)}e^{-i\phi_{i+1}}},
    \label{eq:andreev-crystal-recursion-r-he-ll}
    \\
    r_{\mathrm{eh},LL}^{(i+1)} &= r_{\mathrm{eh},LL}^{(i)} + \frac{r_{\mathrm{eh}}^{(0)} e^{i\phi_{i+1}} t_{\mathrm{ee},LR}^{(i)} t_{\mathrm{hh},LR}^{(i)}}{1 - r_{\mathrm{he},LL}^{(i)} r_{\mathrm{eh}}^{(0)}e^{i\phi_{i+1}}},
    \label{eq:andreev-crystal-recursion-r-eh-ll}
    \\
    t_{\mathrm{ee},LR}^{(i+1)} &= \frac{t_{\mathrm{ee},LR}^{(i)} t_{\mathrm{ee}}^{(0)}}{1 - r_{\mathrm{eh},LL}^{(i)} r_{\mathrm{he}}^{(0)}e^{-i\phi_{i+1}}},
    \label{eq:andreev-crystal-recursion-t-ee-lr}
    \\
    t_{\mathrm{hh},LR}^{(i+1)} &= \frac{t_{\mathrm{hh},LR}^{(i)} t_{\mathrm{hh}}^{(0)}}{1 - r_{\mathrm{he},LL}^{(i)} r_{\mathrm{eh}}^{(0)}e^{i\phi_{i+1}}},
    \label{eq:andreev-crystal-recursion-t-hh-lr}
\end{align}
\end{subequations}
where $r_{\mathrm{eh}}^{(0)} = r_{\mathrm{eh}} e^{i\phi_0}$, $r_{\mathrm{he}}^{(0)} = r_{\mathrm{he}} e^{-i\phi_0}$, $t_{\mathrm{ee}}^{(0)} = t_{\mathrm{ee}}$, and $t_{\mathrm{hh}}^{(0)} = t_{\mathrm{hh}}$. Similar expressions hold for the $RR$ and $RL$ amplitudes.

We calculate the transport properties of a crystal composed of $21$ superconducting leads, corresponding to an array of 20 Josephson junctions with a constant phase difference between neighboring superconductors.
The resulting differential conductance, shown in Fig.~\ref{fig:GLR}(e--h), illustrates how the narrow peaks observed for the NSNSN device in Fig.~\ref{fig:GLR}(a--d) merge into a continuous band, as expected for an extended periodic structure.
A finite electronic temperature of $30\,\mathrm{mK}$, included through Eq.~\eqref{eq:conductance-finite-T}, smooths the otherwise discrete series of sharp resonances associated with the Andreev bound states at zero temperature.

\begin{figure*}
    \centering
    \includegraphics{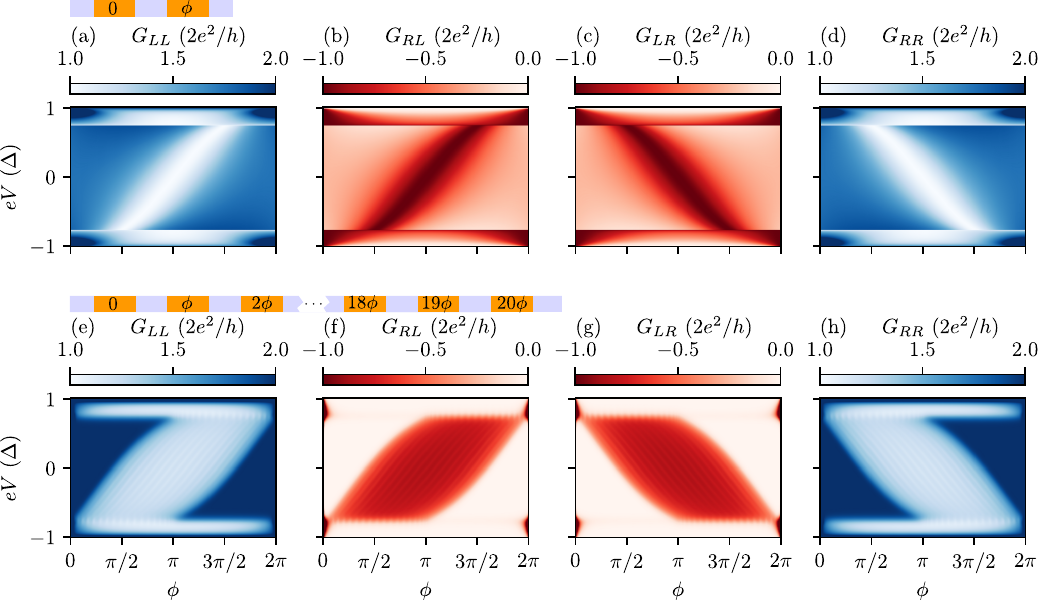}
    \caption{Differential conductance as in Fig.~\ref{fig:GLR}, but with weaker semiconductor--superconductor coupling, $\Gamma = 2\Delta_0$ instead of $6\Delta_0$, which produces a visible minigap. Top panels (a--d): Differential conductance, $G_{\alpha\beta}$, of an NSNSN device with phase difference $\phi$ between the superconductors and bias $eV$. Bottom panels (e--h): Same quantity for a device with 21 superconductors and phase difference $\phi$ between neighboring superconductors. In both cases, the sketch is shown at the top left; panels (a,d) and (e,h) show the local conductance, $\alpha=\beta$, while panels (b,c) and (f,g) show the nonlocal conductance, $\alpha\neq\beta$.}
    \label{fig:GLR_gamma}
\end{figure*}

\subsection{Coupling-induced squeezing}
\label{sec:coupling-induced-squeezing}

We now examine how the proximity effect, incorporated through the self-energy renormalization described in Sec.~\ref{sec:proximity-self-energy}, modifies the results by introducing an explicit dependence on the coupling strength $\Gamma$.

Below the minigap, the Green function in Eq.~\eqref{eq:greens-function} has no real poles, whereas for $\omega$ above the minigap it does. The boundary between these two regimes depends on $\Gamma$.
The emergence of the continuum is illustrated in Fig.~\ref{fig:GLR_gamma}, obtained using the same parameters as Fig.~\ref{fig:GLR} but with a smaller coupling, $\Gamma = 2\Delta_0$ instead of $6\Delta_0$.
This reduced coupling narrows the energy window of the directional conductance.

An analytical estimate of the squeezing effect can be obtained in the NSNSN case by starting from Eq.~\eqref{eq:nsnsn-lorentzian-energy} and including the energy renormalization induced by finite coupling to the superconducting leads. Eqs.~\eqref{eq:x-replacement} and~\eqref{eq:renormalized-penetration-length} can be used to express the directional-conductance bandwidth directly in terms of the microscopic junction parameters, yielding
\begin{equation}
\begin{split}
    \Delta V_{\mathrm{dir}}
    &\approx
    \left(\frac{\dd \omega}{\dd x}\right)_{\!\omega = 0}
    \csch^2\!\left(\frac{\ell_S}{|q^{-1}(\omega = 0)|}\right)
    \\ &\approx
    \frac{\Delta_0}{1+\frac{\Delta_0}{\Gamma}}\,
    \csch^2\!\left(\frac{\ell_S\,\Gamma}{\hbar v_\mathrm{F}}\right).
\end{split}
\label{eq:directional-window}
\end{equation}
Eq.~\eqref{eq:directional-window} provides an analytic estimate of the energy, or bias-voltage, window over which directional transmission occurs.
The first prefactor accounts for the renormalization of quasiparticle energies caused by hybridization between the semiconductor and the superconducting leads, effectively shrinking the accessible energy range as $\Gamma$ decreases. The second term reflects the exponential suppression of coherent transport with increasing superconducting-segment length $\ell_S$. Together, these two factors capture the narrowing of the directional-transmission window observed numerically in Fig.~\ref{fig:GLR}. Appendix~\ref{sec:appendix-coupling-dependence} presents a numerical calculation of how the conductance depends on the semiconductor--superconductor coupling $\Gamma$.

\begin{figure}[ht]
    \centering
    \includegraphics{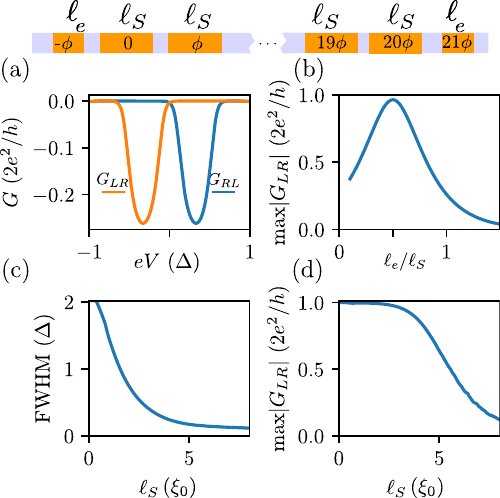}
    \caption{
    (a) Line cuts of Figs.~\ref{fig:GLR}(f,g) at $\phi = 5\pi/4$ ($G_{RL}$ in blue, $G_{LR}$ in orange), showing that, for the same phase bias, a band appears in one direction but not in the other. (b) Adding extra superconductors at the two ends enhances the band, with the peak approaching the conductance quantum, $G_0$. (c,d) Dependence on the length of the central superconductors, with the outer superconductors fixed to $\ell_e = \ell_S/2$, which maximizes the peak. Panel (c) shows the band full width at half maximum, and panel (d) its peak height. Both decrease with increasing superconducting length.
    }
    \label{fig:G1d}
\end{figure}

In Fig.~\ref{fig:G1d}(a), we show a line cut at phase $\phi = 5\pi/4$, also highlighted in Fig.~\ref{fig:GLR}(f,g). At this phase, the nonlocal conductance reaches only about $25\%$ of the conductance quantum, $G_0$. To improve control over the coupling between the leads and the Andreev band, we add two extra superconducting fingers of length $\ell_e$ at the ends of the array, which can be controlled independently of the inner superconducting segments of length $\ell_{S}$. The resulting control over the coupling and broadening via $\ell_e$ is shown in Fig.~\ref{fig:G1d}(b). In particular, when $\ell_e \simeq \tfrac{\ell_{S}}{2}$, the nonlocal conductance approaches $G_0$.

Fig.~\ref{fig:G1d}(c,d) shows how the bandwidth, measured through the full width at half maximum, and the maximum nonlocal conductance are controlled by $\ell_{S}$. As expected, increasing $\ell_{S}$ reduces the bandwidth because the hybridization between Andreev bound states in neighboring junctions becomes weaker. This reduced hybridization also lowers the maximum conductance.

\begin{figure}[ht]
    \centering
    \includegraphics{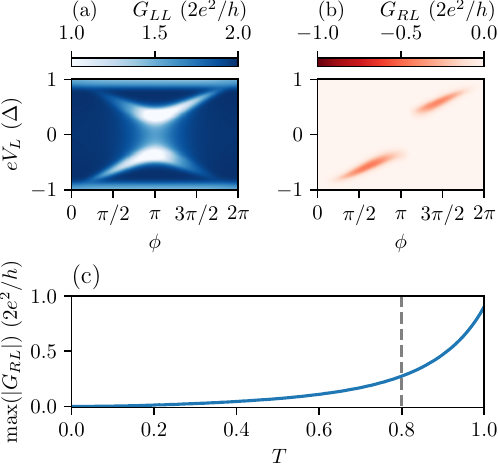}
    \caption{
Numerical simulation of a device with 20 junctions, where the two outer superconductors have length $\ell_S/2$ and the inner superconductors have length $\ell_S = 3\xi_0$. Panels (a) and (b) show the local and nonlocal differential conductance, respectively, illustrating the opening of a gap. The simulations in panels (a) and (b) are performed with transparency $T = 0.8$. Panel (c) shows the effect of varying the transparency, demonstrating that the peak height of the band decreases as $T$ is reduced.}
    \label{fig:finiteT}
\end{figure}

\begin{figure}[hb]
    \centering
    \includegraphics{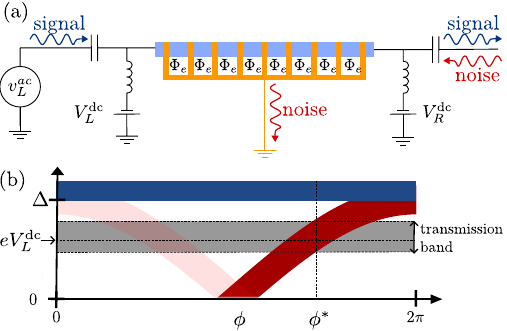}
    \caption{(a) Circuit representation of the proposed directional filter based on an Andreev crystal.
    Static voltages $V^{\mathrm{dc}}_{L}$ and $V^{\mathrm{dc}}_{R}$, together with a uniform phase bias $\phi^{\ast}$, set the operating point.
    An ac drive $v^{\mathrm{ac}}_{L}$ injected at the left port (blue) is transmitted, whereas an identical excitation incident from the right (red) is rejected.
    (b) Sketch of the differential transconductances $G_{RL}$ (dark band) and $G_{LR}$ (light band) as functions of $\phi$ at fixed $V^{\mathrm{dc}}_{L}$.
    Within the shaded window only $G_{RL} \neq 0$; the device therefore operates as a unidirectional link.
    The center and width of this transmission band are set by the combination of $V^{\mathrm{dc}}_{L}$ and $\phi^{\ast}$.}
    \label{fig:filter}
\end{figure}

For imperfect transparency, $T<1$, we numerically concatenate the scattering matrices following Appendix~\ref{sec:appendix-concatenation}. In this regime, normal reflection occurs at the superconductor--normal interfaces, mixing right- and left-movers and opening a gap~\cite{beenakkerUniversalLimitCriticalcurrent1991}. Figs.~\ref{fig:finiteT}(a,b) display the local and nonlocal conductance for $T = 0.8$, showing that a gap opens at $\phi=\pi$. As $T$ decreases, the gap widens and the directional response weakens, reflecting the increased mixing between oppositely moving quasiparticles.
The overall amplitude of the conductance band also decreases with decreasing transparency, as shown in Fig.~\ref{fig:finiteT}(c).

\subsection{Time-dependent transport properties}
\label{sec:time-dependent-transport-properties}

The directional behavior of the differential conductance suggests that the device can function as a directional filter. As illustrated in Fig.~\ref{fig:filter}, we consider a configuration in which a dc voltage bias, $V^{\rm dc}_{L/R}$, is applied independently to each terminal, together with a superimposed sinusoidal ac component. By tuning to an appropriate phase bias, $\phi^*$, the signal is transmitted only in one direction, while noise originating on the right is prevented from reaching the isolated part of the circuit.

To quantify the device response, consider generic time-dependent biases
\begin{equation}
    v_\beta(t) = V_\beta^{\mathrm{dc}} + v_\beta^{\mathrm{ac}}\cos(\omega_D t),
    \label{eq:time-dependent-bias}
\end{equation}
which can be engineered in a setup with bias tees, as shown in Fig.~\ref{fig:filter}(a).
Assuming the adiabatic limit, in which $\hbar\omega_D$ is small compared with the relevant electronic energy scales, we expand the current up to second order in $v_\beta^{\mathrm{ac}}$; see Appendix~\ref{sec:appendix-harmonic-expansion} and Ref.~\cite{pedersenScatteringTheoryPhotonassisted1998}.
The first and second harmonics of the current then read
\begin{align}
    I_{\alpha}^{(|1|)}(t) &=
    -\cos(\omega_D t)
    \sum_{\gamma}
    G_{\alpha\gamma}(V_\gamma^{\mathrm{dc}})\,
    v_\gamma^{\mathrm{ac}},
    \label{eq:first-harmonic-current}
    \\
    I_{\alpha}^{(|2|)}(t) &\approx
    \tfrac{1}{2}\cos(2\omega_D t)
    \sum_{\gamma}
    (v_\gamma^{\mathrm{ac}})^2
    \frac{\partial G_{\alpha\gamma}(V_\gamma^{\mathrm{dc}})}{\partial V_\gamma^{\mathrm{dc}}}.
    \label{eq:second-harmonic-current}
\end{align}

Eq.~\eqref{eq:first-harmonic-current} describes the linear response at the drive frequency $\omega_D$, proportional to both the dc conductance and the applied ac bias, and therefore determines the signal attenuation.
Eq.~\eqref{eq:second-harmonic-current} captures the weak nonlinear correction at $2\omega_D$, which depends on the slope of the conductance curve and provides a measure of harmonic distortion generated by the device.
This distortion can be minimized by adjusting the phase and dc bias within the directional-conductance window.

Therefore, the device can be used as a directional filter, as shown in Fig.~\ref{fig:filter}(b). According to Eq.~\eqref{eq:first-harmonic-current}, the resulting current at the right terminal oscillates at the same frequency with amplitude $I_R^{(|1|)} \propto G_{RL}\,v_L^{\mathrm{ac}}$, producing the transmitted blue signal.
In contrast, when the same ac drive is applied from the right, the corresponding amplitude $I_L^{(|1|)} \propto G_{LR}\,v_R^{\mathrm{ac}}$ is strongly suppressed because $G_{LR}\approx0$ in the directional-conductance regime.
The narrow window in panel~(b), where $G_{RL}\neq0$ but $G_{LR}=0$, therefore defines the operational bandwidth of the Andreev-crystal filter.

\section{Discussion and summary}
\label{sec:discussion-summary}

We have described the transport properties of an Andreev crystal formed in an array of uniformly phase-biased Josephson junctions. Our proposed device consists of a semiconducting channel with an array of superconducting contacts, each with an individually controlled phase. The phase bias can be imposed either by (i) threading a magnetic flux through individual loops that connect neighboring superconductors, or by (ii) connecting each segment to a current-carrying superconducting line that sustains a phase gradient~\cite{lykkegaardPhaseBiasedAndreevDiffraction2025}. When each unit cell carries a phase bias, an array of Andreev bound states is created and the resulting band can be tuned by that phase. For nearly perfect contacts between the normal and superconducting segments, the transport becomes directional because the two branches of the spectrum are then composed purely of right- or left-moving electronic states, respectively.
Backscattering, for example due to imperfect interface transparency, hybridizes the states originating from left- and right-moving electronic modes and thereby opens a gap in the spectrum. For intermediate transparency, the bands retain their directional character away from the gap opening. This design can be extended naturally to two-dimensional crystals, where the relative phase bias along the two directions may differ.

We also point out that an Andreev crystal with constant phase bias can be regarded as a discretized version of a superconductor with finite-momentum Cooper pairing~\cite{fuldeSuperconductivityStrongSpinExchange1964, larkinNONUNIFORMSTATESUPERCONDUCTORS1964, gorkovSuperconducting2DSystem2001, yuanTopologicalMetalsFinitemomentum2021, yuanSupercurrentDiodeEffect2022, daidoIntrinsicSuperconductingDiode2022, hePhenomenologicalTheorySuperconductor2022, ilicTheorySupercurrentDiode2022, palJosephsonDiodeEffect2022, houUbiquitousSuperconductingDiode2023, banerjeePhaseAsymmetryAndreev2023, davydovaUniversalJosephsonDiode2022, davydovaNonreciprocalSuperconductivity2024}.
In a finite-momentum superconductor, the condensate carries a net supercurrent, and the spectrum of quasiparticles moving with and against that current is Doppler shifted~\cite{davydovaUniversalJosephsonDiode2022}.
Intuitively, electrons moving with and against the condensate have different kinetic energies in the condensate frame, and consequently all properties associated with the condensate, including Andreev reflection and the quasiparticle spectrum, are Doppler shifted~\cite{davydovaNonreciprocalSuperconductivity2024}.
As a consequence, quasiparticle modes that co- and counterpropagate with the supercurrent flow are transmissive at different energies.
This gives rise to directional conductance for electrons tunneling through a superconductor that carries supercurrent, whether in a finite-momentum superconductor or in an Andreev crystal where a constant phase bias between superconducting segments induces a net Josephson current.

This direction-selective behavior is a generic feature of Andreev bound states and Andreev crystals. It may serve as a diagnostic probe of normal--superconductor interfaces because of the sensitivity of the interference to backscattering. Finally, we have shown how this behavior can be used to realize a unidirectional filter, allowing a signal to be transmitted in one direction while noise is prevented from propagating in the reverse direction.

The directional transport emerging in Andreev crystals enables new applications for signal processing in superconducting circuits~\cite{ingla-aynesEfficientSuperconductingDiodes2025, krantzQuantumEngineersGuide2019a}.
First, it provides a fully superconducting analogue of a microwave isolator, enabling unidirectional transmission of ac signals between circuit nodes without relying on magnetic materials or ferrites~\cite{Abdo_2014_Josephson}.
Such a device could protect sensitive qubit or detector elements from back-propagating noise while maintaining near-zero dissipation.
Second, because the transmission direction and bandwidth are tunable via the phase bias and bias voltage, the same architecture could function as a reconfigurable on-chip filter or router for superconducting logic and readout lines.
Finally, the intrinsic phase sensitivity of the Andreev modes makes the array attractive as a building block for interferometric sensors or flux-controlled amplifiers.

Hybrid superconductor-semiconductor platforms still lack a defining application beyond their potential for realizing topological states.
The ability to engineer robust, flux- and voltage-tunable nonreciprocal transport directly within such systems could provide precisely that missing functionality.

\section{Acknowledgments}
\label{sec:acknowledgments}

We acknowledge discussions with Magnus R. Lykkegaard.
This research was funded in part by the European Research Council (Grant Agreement No. 856526) and by the DFG Collaborative Research Center (CRC) 183 Project No. 277101999.
M.G. acknowledges support from the German Research Foundation under the Walter Benjamin program (Grant Agreement No. 526129603) and from the Air Force Office of Scientific Research under award number FA2386-24-1-4043.
A.M. and J.S. acknowledge support from the U.S. Office of Naval Research (ONR) through the MURI program (grant U.S. ONR MURI N000142212764).
A.M. also acknowledges funding from the Wallenberg Initiative on Networks and Quantum Information (WINQ).

\section{Data availability}
\label{sec:data-availability}

The code used to generate the data and figures is available on Zenodo~\cite{dahlAndreevCrystalsHybrid2025}.

\appendix

\section{Scattering matrix description}
\label{sec:appendix-scattering-matrix-description}

In general, a scattering matrix $S$ for a scatterer is a unitary matrix defined through
\begin{equation}
    \psi_i^{(o)} = S_{ij} \psi_j^{(i)},
    \label{eq:appendix-scattering-matrix-definition}
\end{equation}
where $\psi_i^{(o/i)}$ is the outgoing or incoming amplitude in channel $i$.

\begin{widetext}
\subsection{Concatenation of scattering matrices}
\label{sec:appendix-concatenation}

Consider two scatterers with scattering matrices $A$ and $B$ and, respectively, $N_A$ and $N_B$ modes. We wish to write the scattering matrix of the composed system $C$, obtained by connecting the last $N_j$ modes of $A$ to the first $N_j$ modes of $B$. A straightforward calculation yields the Redheffer star product~\cite{Redheffer_1962_Relation},
\begin{equation}
    C = A \star_{N_j} B
=
\begin{pmatrix}
    A_{11} + A_{12} (I - B_{11} A_{22})^{-1} B_{11} A_{21} &
    A_{12} (I - B_{11} A_{22})^{-1} B_{12}
    \\
    B_{21} (I - A_{22} B_{11})^{-1} A_{21} &
    B_{22} + B_{21} (I - A_{22} B_{11})^{-1} A_{22} B_{12}
\end{pmatrix}.
\label{eq:redheffer-star-product}
\end{equation}
\end{widetext}
The resulting matrix $C$ has dimension $N_A + N_B - 2N_j$, while the subblocks $B_{11}$ and $A_{22}$, as well as the identity matrix $I$, are $N_j\times N_j$. The remaining dimensions are inferred from context. The product is well defined provided $(I - B_{11} A_{22})$ and $(I - A_{22} B_{11})$ are invertible, which physically means that no bound states are formed between the two scatterers.

Consider next a system with scattering matrix $S$, and suppose that open or periodic boundary conditions are imposed on a subset of the leads. Separating the channels subject to boundary conditions from the open channels, the scattering problem can be written in block form as
\begin{equation}
    \begin{pmatrix}
        \psi_d^{(o)} \\
        \psi_l^{(o)}
    \end{pmatrix}
    =
    \begin{pmatrix}
        S_{dd} & S_{dl} \\
        S_{ld} & S_{ll}
    \end{pmatrix}
    \begin{pmatrix}
        \psi_d^{(i)} \\
        \psi_l^{(i)}
    \end{pmatrix},
    \label{eq:block-scattering-problem}
\end{equation}
where $l$ denotes the subblock of open channels and $d$ denotes the subblock of channels subject to boundary conditions. These conditions can be written as a linear relation of the form
\begin{equation}
     \psi^{(i)}_d = \Lambda \psi^{(o)}_d,
     \label{eq:boundary-condition-matrix}
\end{equation}
where the matrix $\Lambda$ encodes the boundary conditions. One then obtains a closed equation for the scattering matrix of the open leads,
\begin{equation}
     \tilde{S}_{ll} = S_{ll} + S_{ld} (I - \Lambda S_{dd})^{-1} \Lambda S_{dl}.
     \label{eq:effective-open-scattering-matrix}
\end{equation}

\subsection{Normal region scattering matrix}
\label{sec:appendix-normal-region-scattering-matrix}

For the calculation, we work within the Andreev approximation and linearize the dispersion relation in the vicinity of the Fermi surface. In this way, the normal-state quasiparticle excitation energy is $\xi_k = \hbar \tilde{v}_\mathrm{F} q$, where $\tilde{v}_\mathrm{F}$ is the Fermi velocity in the normal region. We can invert the dispersion relation in the normal region as
\begin{equation}
    \tilde{k} = \tilde{k}_\mathrm{F} + q \simeq \tilde{k}_{\mathrm{F}} + \tau\frac{\xi_k}{\hbar \tilde{v}_{\mathrm{F}}},
    \label{eq:normal-region-wavevector}
\end{equation}
where $\tau=\pm1$ denotes the particle-hole quantum number, while in the superconducting region we have
\begin{equation}
    k = k_\mathrm{F} + \tau \frac{\sqrt{\omega^2 - \Delta^2}}{\hbar v_\mathrm{F}} = k_\mathrm{F} + \frac{\tau}{\pi\xi_0} \sqrt{\frac{\omega^2}{\Delta^2} - 1},
    \label{eq:superconducting-region-wavevector}
\end{equation}
with
\begin{equation}
    \xi_0^{-1} = \frac{\pi\Delta}{\hbar v_\mathrm{F}}.
    \label{eq:appendix-bare-coherence-length}
\end{equation}
When $\omega<\Delta$, the wave vectors become complex, corresponding to evanescent waves.

We begin with the normal region.
The scattering matrix for a perfectly clean metal reads
\begin{equation}
S_\mathrm{N, cln} =
\begin{pmatrix}
    0 & 0 & e^{+i k_e \ell_\mathrm{N}} & 0 \\
    0 & 0 & 0 & e^{-i k_h \ell_\mathrm{N}} \\
    e^{+i k_e \ell_\mathrm{N}} & 0 & 0 & 0 \\
    0 & e^{-i k_h \ell_\mathrm{N}} & 0 & 0 \\
\end{pmatrix},
\label{eq:clean-normal-scattering-matrix}
\end{equation}
where $\ell_\mathrm{N}$ is the length of the normal region. In the case of a disordered medium, we parametrize the scattering matrix as
\begin{equation}
    S_\mathrm{N, dis} = \begin{pmatrix}
    +r & t\\
    t & -r
\end{pmatrix}  \otimes \tau_0 =
\begin{pmatrix}
    +r & 0 & t & 0\\
    0 & +r & 0 & t  \\
    t & 0 & -r & 0 \\
    0 & t & 0 & -r
\end{pmatrix}.
\label{eq:disordered-normal-scattering-matrix}
\end{equation}
Here $t$ and $r$ are the transmission and reflection amplitudes. We parametrize them by a single transparency $T\in[0,1]$, defined as the transmission probability $T\equiv |t|^2$. In the short-junction limit, these can be approximated as energy-independent real amplitudes,
\begin{equation}
    r=\sqrt{1-T},
    \qquad
    t=\sqrt{T},
    \label{eq:appendix-transparency-parametrization}
\end{equation}
as discussed around Eq.~\eqref{eq:nsn-amplitudes}. The matrix $S_\mathrm{N}$ is unitary and satisfies particle-hole symmetry. As in Ref.~\cite{beenakkerUniversalLimitCriticalcurrent1991}, we neglect the energy dependence of the normal-region scattering matrix.

\subsection{NSN scattering matrix}
\label{sec:appendix-nsn-scattering-matrix}

Consider a clean one-dimensional metal in which a section of length $\ell_\mathrm{S}$ is superconducting. For an incoming electron from the left, the wave function in the three regions can be written as
\begin{equation}
    \begin{cases}
    \psi_{I} = \begin{pmatrix}
    1 \\ 0 \end{pmatrix} e^{i \tilde{k}_e z} + a \begin{pmatrix}
    0 \\ 1 \end{pmatrix} e^{+ i \tilde{k}_h z}, \\
    \psi_{II} = b \begin{pmatrix}
    u \\ v \end{pmatrix} e^{i k_e z} + d \begin{pmatrix}
    v \\ u \end{pmatrix} e^{+ i k_h z}, \\
    \psi_{III} = f \begin{pmatrix}
    1 \\ 0 \end{pmatrix} e^{i \tilde{k}_e (z - \ell_\mathrm{S})},
    \end{cases}
    \label{eq:appendix-nsn-wavefunctions}
\end{equation}
where $z$ is the longitudinal coordinate, $\tilde{k}_e$ and $\tilde{k}_h$ are the wave vectors in the normal metal, while $k_e = k_\mathrm{F}+q$ and $k_h = k_\mathrm{F}-q$ are the corresponding wave vectors in the superconducting segment.

By wave-function matching, we obtain the linear system
\begin{equation}
    \begin{cases}
        1 = b u + d v, \\
        a = b v + d u, \\
        b u e^{i k_e \ell_\mathrm{S}} + d v e^{+ i k_h \ell_\mathrm{S}} = f,\\
        b v e^{i k_e \ell_\mathrm{S}} + d u e^{+ i k_h \ell_\mathrm{S}} = 0,
    \end{cases}
    \label{eq:appendix-nsn-linear-system}
\end{equation}
which gives the following solutions
\begin{align}
    r_{he} &= a = \frac{(e^{-i q \ell_\mathrm{S}} - e^{+i q \ell_\mathrm{S}}) u v}{e^{-i q \ell_\mathrm{S}} u^2 - e^{+i q \ell_\mathrm{S}} v^2},
    \label{eq:appendix-nsn-r-he-solution}
    \\
    t_{ee} &= f = \frac{u^2 - v^2}{e^{-i q \ell_\mathrm{S}} u^2 - e^{+i q \ell_\mathrm{S}} v^2}  e^{+i k_F \ell_\mathrm{S}}.
    \label{eq:appendix-nsn-t-ee-solution}
\end{align}

After some algebra, and defining the infinite-length-limit Andreev-reflection amplitude
\begin{equation}
    r_{he, \Delta} = \frac{v}{u} = \exp(-i \arccos(E/\Delta)),
    \label{eq:appendix-infinite-length-andreev-amplitude}
\end{equation}
and the high-energy-limit normal-transmission amplitude
\begin{equation}
    t_{ee, \mathrm{HE}} =  \exp(i \frac{\ell_\mathrm{S}}{\xi_0} \sqrt{E^2/\Delta^2-1}),
    \label{eq:appendix-high-energy-transmission-amplitude}
\end{equation}
we obtain the final form
\begin{align}
    r_{he} &= \frac{1 - t_{ee, \mathrm{HE}}^2}{1 - t_{ee, \mathrm{HE}}^2 r_{he, \Delta}^2} r_{he, \Delta},
    \label{eq:appendix-final-r-he}
    \\
    t_{ee} &= \frac{1 - r_{he, \Delta}^2}{1 - t_{ee, \mathrm{HE}}^2 r_{he, \Delta}^2}  t_{ee, \mathrm{HE}}  e^{+i k_F \ell_\mathrm{S}}.
    \label{eq:appendix-final-t-ee}
\end{align}

In the long-$\ell_\mathrm{S}$ limit, these simplify to
\begin{align}
    r_{he} &\simeq (1 - t_{ee, \mathrm{HE}}^2) r_{he, \Delta},
    \label{eq:appendix-long-segment-r-he}
    \\
    t_{ee} &\simeq t_{ee, \mathrm{HE}}  e^{+i k_F \ell_\mathrm{S}},
    \label{eq:appendix-long-segment-t-ee}
\end{align}
while preserving the unitarity of the scattering matrix. The final result for the superconducting-segment scattering matrix is
\begin{equation}
    S_\mathrm{S} = \begin{pmatrix}
        0 & r_{eh} & t_{ee} & 0 \\
        r_{he} & 0 & 0 & t_{hh} \\
        t_{ee} & 0 & 0 & r_{eh} \\
        0 & t_{hh} & r_{he} & 0 \\
    \end{pmatrix},
    \label{eq:appendix-final-superconducting-scattering-matrix}
\end{equation}
where $t_{hh}(\omega) = t_{ee}^*(-\omega)$ and
$r_{eh}(\omega) = -r_{he}^*(-\omega)$ because of particle-hole symmetry.

Although these expressions were derived in the clean limit, they are written in terms of universal quantities such as the gap parameter and coherence length. We therefore use them as a phenomenological parametrization when constructing the scattering matrix of the composite system, assigning all the normal reflection to the normal-region scattering matrix.

\section{Landauer--Büttiker approach}
\label{sec:appendix-landauer-buttiker}

Within the Landauer--Büttiker framework, the current in a multiterminal device can be expressed in terms of transmission probabilities derived from the scattering matrix. For a superconducting hybrid structure, the scattering matrix has Nambu structure, and we denote by
\begin{equation}
    T_{\alpha \beta}^{\gamma \delta} (\omega) \equiv \tr \abs{S_{\alpha \beta}^{\gamma \delta}(\omega)}^2,
    \label{eq:transport-probability-definition}
\end{equation}
the probability for a quasiparticle of type $\delta \in \{e,h\}$ entering from lead $\beta$ to exit as type $\gamma \in \{e,h\}$ in lead $\alpha$ at energy $\omega$.

\begin{widetext}
In the superconducting version of the Landauer--Büttiker theory, quasiparticle conservation replaces electron conservation in conventional devices. As a consequence, electric charge is not explicitly conserved. One can distinguish between the charge-conserving normal processes $T^{ee}_{\alpha \beta}$ and $T^{hh}_{\alpha \beta}$, and the charge-converting Andreev processes $T^{eh}_{\alpha \beta}$ and $T^{he}_{\alpha \beta}$. The latter two correspond to the creation or destruction of one Cooper pair in the superconducting leads, respectively~\cite{Lesovik_PU_2011, danonNonlocalConductanceSpectroscopy2020, maianiConductanceMatrixSymmetries2022}.

The average current flowing into lead $\alpha$ in the spin-degenerate case reads
\begin{equation}
\begin{split}
I_\alpha =  & \frac{2 e}{h} \int_{-\infty}^{+\infty}\dd{\omega}\, \left[f(\omega - e V_{\alpha },  \theta) - f(\omega, \theta)\right] \left[N_{\alpha} - R_\alpha^{ee}(\omega) + R_\alpha^{he}(\omega) \right] \\
- & \sum_{\beta\neq\alpha} \frac{2 e}{h} \int_{-\infty}^{+\infty}\dd{\omega}\, \left[f(\omega - e V_{\beta},  \theta) - f(\omega, \theta)\right] \left[T_{\alpha \beta}^{ee}(\omega) -  T_{\alpha \beta}^{he}(\omega)\right],
\end{split}
\label{eq:appendix-average-current}
\end{equation}
where $f(\omega, \theta) = [1+\exp(\omega/k_B \theta)]^{-1}$ is the Fermi--Dirac distribution, $V_\alpha$ are the voltage biases measured with respect to ground, and $\theta$ is the electron temperature.
\end{widetext}

The differential conductance follows directly as $G_{\alpha \beta} = \dd I_\alpha / \dd V_{\beta}$, resulting in
\begin{equation}
\begin{split}
G_{\alpha \alpha} &= G_0 \int_{-\infty}^{+\infty} \dd{\omega}\, h(\omega - e V_{\alpha}) [N_{\alpha} - R_\alpha^{ee}(\omega) + R_\alpha^{he}(\omega)],
\end{split}
\label{eq:appendix-local-differential-conductance}
\end{equation}
for the local differential conductance, while for the nonlocal differential conductance we obtain
\begin{equation}
\begin{split}
G_{\alpha \beta}& = - G_0 \int_{-\infty}^{+\infty}\dd{\omega}\, h(\omega - e V_{\beta})[T_{\alpha \beta}^{ee}(\omega) -  T_{\alpha \beta}^{he}(\omega)],
\end{split}
\label{eq:appendix-nonlocal-differential-conductance}
\end{equation}
where $G_0=\frac{2 e^2}{h}$ is the conductance quantum and
\begin{equation}
    h(\omega, \theta) = - \partial_{\omega} f(\omega, \theta) = \frac{1}{2 k_B \theta} \frac{1}{1 + \cosh (\omega/k_B \theta)}
\label{eq:appendix-thermal-broadening-kernel}
\end{equation}
is the thermal-broadening kernel.

\section{Harmonic expansion of the ac current response}
\label{sec:appendix-harmonic-expansion}

In this appendix, we derive the harmonic content of the time-dependent current within the scattering formalism of Pedersen and Büttiker~\cite{pedersenScatteringTheoryPhotonassisted1998}.
We consider a multiterminal hybrid device subject to both dc and ac voltage drives. Each terminal $\gamma$ is biased by a time-dependent voltage 
\begin{equation}
    V_\gamma(t) = V_\gamma^{\mathrm{dc}} + v_\gamma^{\mathrm{ac}}\cos(\omega_D t),
\end{equation}
where $V_\gamma^{\mathrm{dc}}$ is the static bias, $v_\gamma^{\mathrm{ac}}$ the ac amplitude, and $\omega_D$ the drive frequency.

\begin{widetext}
The current entering lead $\alpha$ can be expressed in terms of the scattering matrix through the generalized formula
\begin{align}
    I_{\alpha}(t) = \frac{e}{h} \int dE \sum_{\gamma,l,k} \Tr \qty[ A_{\gamma\gamma} (\alpha, E, E+ (k-l)\hbar\omega_D) ] J_l\qty(\frac{ev_\gamma^{\mathrm{ac}}}{\hbar\omega_D}) J_k\qty(\frac{ev_\gamma^{\mathrm{ac}}}{\hbar\omega_D}) e^{-i(k-l)\omega_D t} f(E - e V_\gamma^{\text{dc}} - l\hbar\omega_D),
\label{eq:pedersen-formula}
\end{align}
where $e$ is the electron charge and $h$ Planck's constant, $f(E) = [1+\exp(E/k_{\rm B} \theta)]^{-1}$ is the Fermi distribution at temperature $\theta$, $A_{\gamma\gamma}(\alpha,E,E')$ is the scattering kernel describing transmission from terminal $\gamma$ into $\alpha$ with an energy exchange $E'-E$, $J_l$ are Bessel functions of the first kind arising from the Floquet decomposition of the ac bias, and $l,k \in \mathbb{Z}$ count the number of photons absorbed from or emitted into the drive.
\end{widetext}

We are interested in the regime $\hbar \omega_D \ll \Delta, E_F$, where $\Delta$ is the induced superconducting gap and $E_F$ the Fermi energy. In this limit, the Fermi function in Eq.~\eqref{eq:pedersen-formula} can be expanded in powers of $\omega_D$ as
\begin{equation}
\begin{split}
    f(E - eV_\gamma^{\text{dc}} - l\hbar\omega_D)
    &\approx
    f(E - eV_\gamma^{\text{dc}})
    \\
    &\quad + l\hbar\omega_D f'(E - eV_\gamma^{\text{dc}})
    \\
    &\quad + l^2 \frac{\hbar^2 \omega_D^2}{2} f''(E - eV_\gamma^{\text{dc}})
    + \ldots \, .
\end{split}
\label{eq:appendix-fermi-expansion}
\end{equation}
The linear term generates the first harmonic response, while the quadratic term contributes to the second harmonic.

To obtain the first harmonic, we restrict to processes with $k-l = \pm 1$. For $k-l = 1$, inserting the expansion into Eq.~\eqref{eq:pedersen-formula} gives
\begin{equation}
\begin{split}
    I_{\alpha}^{(1)} \approx & \frac{2e}{h} \int dE \sum_{\gamma,l} \Tr \qty[ A_{\gamma\gamma} (\alpha, E, E+ \hbar\omega_D) ]
    \\
    &\times J_l\qty(\frac{ev_\gamma^{\mathrm{ac}}}{\hbar\omega_D}) J_{l+1}\qty(\frac{ev_\gamma^{\mathrm{ac}}}{\hbar\omega_D}) e^{-i\omega_D t}
    \\
    &\times \qty( f(E - eV_\gamma^{\text{dc}}) - l\hbar\omega_D f'(E - eV_\gamma^{\text{dc}}) ).
\end{split}
\label{eq:appendix-first-harmonic-intermediate}
\end{equation}
Using the Bessel-function identities
\begin{align}
    \sum_l J_l(x) J_{l+n}(x) &= \delta_{n,0},
    \label{eq:appendix-bessel-identity-one}
    \\
    \sum_l l J_l(x) J_{l+n}(x) &= \frac{x}{2} (\delta_{n,1} + \delta_{n,-1}),
    \label{eq:appendix-bessel-identity-two}
\end{align}
the sum over $l$ can be evaluated, leading to
\begin{equation}
\begin{split}
    I_{\alpha}^{(1)}(t) &\approx -\frac{2e}{h} e^{-i\omega_D t}\int dE \sum_{\gamma} \frac{ev_\gamma^{\mathrm{ac}}}{2}
    \\
    &\times \Tr \qty[ A_{\gamma\gamma} (\alpha, E, E+ \hbar\omega_D) ] \frac{\partial}{\partial E}f(E - eV_\gamma^{\text{dc}}).
\end{split}
\label{eq:appendix-first-harmonic-after-bessel}
\end{equation}

For $\omega_D \to 0$ we approximate $A_{\gamma\gamma}(\alpha,E,E+\hbar\omega_D)\simeq A_{\gamma\gamma}(\alpha,E,E)$. Introducing the differential conductance
\begin{equation}
    G_{\alpha\gamma} = \frac{2 e^2}{h} \int \mathrm{d} E \Tr \qty[ A_{\gamma\gamma} (\alpha, E, E) ] \frac{\partial }{\partial E} f(E - eV_\gamma^{\text{dc}}),
    \label{eq:appendix-differential-conductance-definition}
\end{equation}
we obtain
\begin{align}
    I_{\alpha}^{(1)}(t) &\approx -e^{-i\omega_D t} \sum_{\gamma} G_{\alpha\gamma} \frac{v_\gamma^{\mathrm{ac}}}{2}.
    \label{eq:appendix-first-harmonic-positive}
\end{align}
Combining this with the corresponding contribution for $k-l = -1$,
\begin{align}
    I_{\alpha}^{(-1)}(t) &\approx -e^{i\omega_D t} \sum_{\gamma} G_{\alpha\gamma} \frac{v_\gamma^{\mathrm{ac}}}{2},
    \label{eq:appendix-first-harmonic-negative}
\end{align}
we obtain the first-harmonic current
\begin{align}
    I_{\alpha}^{(|1|)}(t) &= -\cos(\omega_D t) \sum_{\gamma} G_{\alpha\gamma} v_\gamma^{\mathrm{ac}}.
    \label{eq:appendix-first-harmonic-final}
\end{align}
This term is linear in the ac drive and proportional to the dc conductance.

\begin{widetext}
The same procedure yields the second harmonic by considering $k-l = \pm 2$. Starting with $k-l = 2$, we have
\begin{align}
    I_{\alpha}^{(2)}(t) \approx \frac{2e}{h} \int dE \sum_{\gamma,l} \Tr \qty[ A_{\gamma\gamma} (\alpha, E, E+ 2\hbar\omega_D) ] J_l\qty(\frac{ev_\gamma^{\mathrm{ac}}}{\hbar\omega_D}) J_{l+2}\qty(\frac{ev_\gamma^{\mathrm{ac}}}{\hbar\omega_D}) e^{-i2\omega_D t}
    \\
    \times \qty( f(E - eV_\gamma^{\text{dc}}) - l\hbar\omega_D f'(E - eV_\gamma^{\text{dc}}) + \frac{l^2 \hbar^2 \omega_D^2}{2} f''(E - eV_\gamma^{\text{dc}}) ).
\label{eq:appendix-second-harmonic-intermediate}
\end{align}
Only the last term contributes to the second harmonic, so
\begin{align}
    I_{\alpha}^{(2)}(t) &\approx \frac{2e}{h} e^{-i2\omega_D t}\int dE \sum_{\gamma} \frac{e^2 \qty(v_\gamma^{\mathrm{ac}})^2}{4} \Tr \qty[ A_{\gamma\gamma} (\alpha, E, E+ 2\hbar\omega_D) ] \frac{\partial^2}{\partial E^2}f(E - eV_\gamma^{\text{dc}}),
    \label{eq:appendix-second-harmonic-after-selection}
\end{align}
where we have used the Bessel identity
\begin{align}
    \sum_l l^2 J_l(x) J_{l+n}(x) = \frac{x}{2}(\delta_{n,-1} - \delta_{n,1} ) + \frac{x^2}{4} (\delta_{n,2} + \delta_{n,-2}).
    \label{eq:appendix-bessel-identity-three}
\end{align}
\end{widetext}

Again, we focus on the low-frequency limit. By simple algebra, we obtain
\begin{align}
    I_{\alpha}^{(2)}(t) &\approx e^{-i2\omega_D t} \sum_\gamma \qty(\frac{v_\gamma^\mathrm{ac}}{2})^2 \frac{\partial G_{\alpha\gamma}}{\partial V_\gamma^\mathrm{dc}},
    \label{eq:appendix-second-harmonic-positive}
\end{align}
and similarly for $k-l = -2$, we find
\begin{align}
    I_{\alpha}^{(-2)}(t) &\approx e^{i2\omega_D t} \sum_\gamma \qty(\frac{v_\gamma^\mathrm{ac}}{2})^2 \frac{\partial G_{\alpha\gamma}}{\partial V_\gamma^\mathrm{dc}},
    \label{eq:appendix-second-harmonic-negative}
\end{align}
so that the second harmonic becomes
\begin{align}
    I_{\alpha}^{(|2|)}(t) &\approx \frac{1}{2} \cos\qty(2\omega_D t) \sum_\gamma \qty(v_\gamma^\mathrm{ac})^2 \frac{\partial G_{\alpha\gamma}}{\partial V_\gamma^\mathrm{dc}}.
    \label{eq:appendix-second-harmonic-final}
\end{align}

\section{Dependence of conductance on coupling $\Gamma$ to the superconducting leads}
\label{sec:appendix-coupling-dependence}

\begin{figure*}
    \centering
    \includegraphics[width=\linewidth]{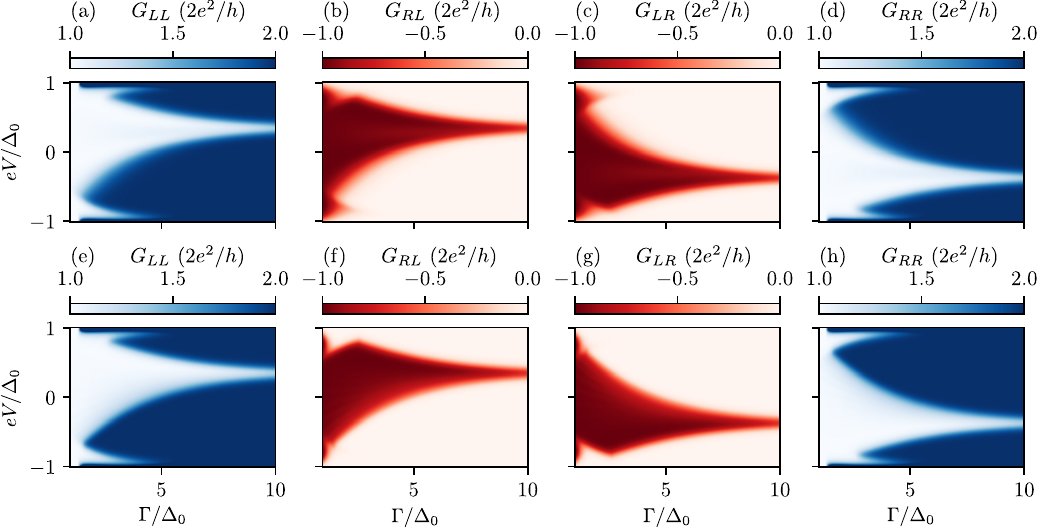}
    \caption{ Local and nonlocal conductance as a function of the coupling strength $\Gamma$ in units of the parent gap $\Delta_0$ for an NSNSN setup (top) and a 20-junction array (bottom). Top panels (a)–(d): Differential conductance, $G_{\alpha\beta}$, of an NSNSN device. Bottom panels (e)–(h): Same quantity for a device with 20 junctions. In both cases, panels (a) and (d) and panels (e) and (h) show the local conductance, $\alpha=\beta$, while panels (b) and (c) and panels (f) and (g) show the nonlocal conductance, $\alpha\neq\beta$. The remaining simulation parameters are $\Delta_0 = \SI{340}{\mu eV}$, $m^* = 0.023m_e$, $\phi = \pi/4$, $\ell_{S} = 3\xi_0$, $\mu = \SI{1}{meV}$, and $\theta = \SI{30}{mK}$.}
    \label{fig:app-GLR-Gamma}
\end{figure*}

We investigated the dependence of the conductance matrix on the coupling strength $\Gamma$ between the semiconductor and the superconducting leads; see Fig.~\ref{fig:app-GLR-Gamma}. Reducing $\Gamma$ broadens the conducting Andreev bands. At $\Gamma = 0$, there is no Andreev reflection, and the conductance reduces to that of a normal semiconductor segment.

\clearpage

\bibliography{QT}

\end{document}